\documentstyle[12pt,epsfig]{article}
\newcommand{\NP}[3]{{Nucl. Phys.}        {#1} {(19#2)} {#3}}

\newcommand{\czdot}{\! \cdot \!}
\newcommand{\beq}{\begin{equation}}
\newcommand{\eeq}{\end{equation}}
\newcommand{\beqa}{\begin{eqnarray}}
\newcommand{\eeqa}{\end{eqnarray}}
\newcommand{\beqan}{\begin{eqnarray*}}
\newcommand{\eeqan}{\end{eqnarray*}}
\newcommand{\ba}{\begin{array}}
\newcommand{\ea}{\end{array}}
\newcommand{\ben}{\begin{enumerate}}
\newcommand{\een}{\end{enumerate}}
\newcommand{\bfl}{\begin{flushleft}}
\newcommand{\efl}{\end{flushleft}}
\newcommand{\btab}{\begin{tabular}}
\newcommand{\etab}{\end{tabular}}
\newcommand{\bit}{\begin{itemize}}
\newcommand{\eit}{\end{itemize}}
\newcommand{\bdes}{\begin{description}}
\newcommand{\edes}{\end{description}}
\newcommand{\bdm}{\begin{displaymath}}
\newcommand{\edm}{\end{displaymath}}

\newcommand{\no}{\nonumber}

\newcommand{\ra}{\rightarrow}

\newcommand{\ve}{\varepsilon}
\newcommand{\vp}{\varphi}

\newcommand{\wt}{\widetilde}

\newcommand{\cL}{{\cal L}}

\begin{document}
\begin{titlepage}
\begin{flushright}
INFNNA-IV-96/11\\
UWThPh-1996-10\\
LNF-96/013(P) \\
March 1996\\
\end{flushright}
\begin{center}
%{\Large DRAFT} \\ [3pt]
%\today
 
\vspace*{1cm}
 
{\large \bf Radiative Four--Meson Amplitudes \\[8pt]
in Chiral Perturbation Theory*} 
 
\vspace*{1cm}
{\bf{G. D'Ambrosio$^1$, G. Ecker$^2$, G. Isidori$^3$ and H. Neufeld$^2$}}
 
\vspace{.5cm}
${}^{1)}$ INFN, Sezione di Napoli \\
Dipartimento di Scienze Fisiche, Universit\`a di Napoli\\
I--80125 Napoli, Italy \\[5pt]
 
${}^{2)}$ Institut f\"ur Theoretische Physik, Universit\"at Wien\\
A--1090 Wien, Austria \\[5pt]
 
${}^{3)}$ INFN, Laboratori Nazionali di Frascati \\ 
P.O. Box 13, I--00044 Frascati, Italy

\vfill
{\bf Abstract} \\
\end{center}
\noindent
We present a general discussion of radiative four--meson processes 
to $O(p^4)$ in chiral perturbation theory. 
We propose a definition of ``generalized bremsstrahlung'' that 
takes full advantage of experimental information on
the corresponding non--radiative process. We also derive general 
formulae for one--loop amplitudes which can be applied, for instance, 
to $\eta \ra 3\pi\gamma$, $\pi \pi \ra \pi \pi \gamma$ and 
$K \ra 3\pi\gamma$.

\vfill
\noindent * Work supported in part
by HCM, EEC--Contract No. CHRX--CT920026 (EURODA$\Phi$NE) and
by FWF (Austria), Project Nos. P09505--PHY, P10876-PHY
\end{titlepage}

\paragraph{1.}  Chiral perturbation theory (CHPT) \cite{Wein79,GL84,GL85}
incorporates electromagnetic gauge invariance. To lowest
order in the derivative expansion, $O(p^2)$ in the meson sector, amplitudes
for radiative transitions are
completely determined by the corresponding non--radiative amplitudes.
Direct emission, carrying genuinely new information, appears only at
$O(p^4)$. Nevertheless, even in higher orders of the chiral expansion part
of the amplitude is related to the respective non--radiative process.
In order to isolate the direct emission amplitude, an
operational definition of ``generalized bremsstrahlung'' is needed.

In this letter, we investigate in a general manner radiative transitions
involving four pseudoscalar mesons and one real photon. Possible
applications to be discussed elsewhere include $\eta \ra 3\pi\gamma$
and $\pi \pi \ra \pi \pi \gamma$ in the strong sector and the
nonleptonic weak decays $K \ra 3\pi\gamma$. Our purpose is twofold:
\begin{enumerate}
\item[i.] We extend Low's theorem \cite{Low58} by terms of $O(k)$ 
($k$ is the photon momentum) to define generalized bremsstrahlung. This 
part will include in particular {\bf all} local
terms of $O(p^4)$ that contribute also to the non--radiative four--meson
transition.
\item[ii.] We give a compact expression for the loop amplitude of a
general four--meson process with a real photon. The resulting formula is
immediately applicable to both strong and nonleptonic weak processes.
We also consider the limiting case of a radiative three--meson amplitude
to recover known results for $K \ra 2\pi\gamma$ decays
\cite{ENP92,DMS93,ENP94,DI95}.
\end{enumerate}
\paragraph{2.} The amplitude for a four--meson transition with a single
photon can be decomposed into an electric and a magnetic part:
\beq
A(\vp_a \vp_b \vp_c \vp_d \gamma) = e \ve^\mu(k)(E_\mu + 
\ve_{\mu\nu\rho\sigma} M^{\nu\rho\sigma})
\eeq
with
$$
k^\mu E_\mu = 0, \qquad \ve_{\mu\nu\rho\sigma} k^\mu M^{\nu \rho\sigma}
= 0~.
$$
The magnetic amplitude $M^{\nu\rho\sigma}$ can only occur in nonleptonic
weak processes (for a review, see Ref.~\cite{DEIN95}). It appears first
at $O(p^4)$ as a tree--level contribution and will not concern us
further. Here, we are only interested in the electric amplitude $E_\mu$
that is in particular sensitive to bremsstrahlung. The kinematics of
the process is specified by five scalar variables which we choose as
\beq
s = (p_1 + p_2)^2, \qquad \nu = p_3(p_1 -p_2), \qquad
t_i = k \czdot p_i \qquad (i = 1,\ldots,4) \label{kinema}
\eeq
with
$$
\sum_{i=1}^4 p_i + k = 0, \qquad 
t_1 + t_2 + t_3 + t_4 = 0~.
$$
Any three of the $t_i$ together with $s$ and $\nu$ form a
set of independent variables.

The non--radiative transition is characterized by the two Dalitz variables
$s$ and $\nu$. 
Denoting the non--radiative amplitude by $A(s,\nu)$, Low's theorem
\cite{Low58} amounts to the following expansion in the photon momentum
$k$:
\beqa
\label{Low}
E^\mu &=& A(s,\nu) \Sigma^\mu \no \\*
&& \mbox{} + 2 \frac{\partial A(s,\nu)}{\partial s} \Lambda^\mu_{12} +
\frac{\partial A(s,\nu)}{\partial \nu} (\Lambda^\mu_{13} -
\Lambda^\mu_{23}) \no \\*
&& \mbox{} + O(k)
\eeqa
with (the meson charges in units of $e$ are denoted $q_i$)
\beqa
\label{defs}
\Sigma^\mu &=& \sum_{i=1}^4 \frac{q_i p_i^\mu}{t_i} \no \\
\Lambda^\mu_{ij} &=& \Lambda^\mu_{ji} = (q_i t_j - q_j t_i) D^\mu_{ij}
\no \\
D^\mu_{ij} &=& - D^\mu_{ji} = \frac{p_i^\mu}{t_i} - \frac{p_j^\mu}{t_j}~.
\eeqa
The explicit terms in (\ref{Low}) are often called ``internal
bremsstrahlung".
It is straightforward to show that there are no terms of $O(k)$ at 
lowest order in the chiral expansion. Thus, for radiative four--meson
processes the leading chiral amplitude of $O(p^2)$ is completely 
determined by the non--radiative amplitude $A(s,\nu)$ as expressed by
Eq.~(\ref{Low}).

\paragraph{3.} At $O(p^4)$, there are as usual both one--loop and 
tree--level contributions with a single vertex from the strong
Lagrangian $\cL_4$ \cite{GL84,GL85} or the nonleptonic weak Lagrangian
$\cL_4^{\Delta S=1}$ \cite{KMW90,EKW93}. Let us first consider the
tree--level amplitude. The different terms in either $\cL_4$ or
$\cL_4^{\Delta S=1}$ that can contribute to the processes under
consideration can be grouped in four classes:
\begin{enumerate}
\item[A.] Terms of $O(m_q^2)$ without derivatives: fully covered
by internal bremsstrahlung (\ref{Low}).
\item[B.] Terms of $O(m_q)$ with two (covariant) derivatives: again
included in (\ref{Low}).
\item[C.] Four--derivative terms: in general not 
fully covered by (\ref{Low}).
\item[D.] Terms with two derivatives and one field strength tensor
containing the electromagnetic field: contribute only to 
the radiative transition and thus are never included in (\ref{Low}).
\end{enumerate}

Obviously, groups A,B correspond to internal bremsstrahlung while the
contributions of type D belong to direct emission. Class C falls in
between if we adopt Eq.~(\ref{Low}) as the definition of
bremsstrahlung. On the other hand, it would have both conceptual and
practical advantages to include {\bf all} terms under the heading
``bremsstrahlung'' that contribute to both radiative and
non--radiative transitions. One practical advantage arises for
$K \ra 3\pi\gamma$ decays where the low--energy constants of the
four--derivative terms are only partly known \cite{KMW91,EKW93}. If those 
terms could be included in bremsstrahlung, we may use experimental data for
$K \ra 3\pi$ decays directly without having to worry about the values of
the aforementioned coupling constants \cite{DEIN962}.

In order to incorporate class C in what we shall call generalized
bremsstrahlung, we must add
explicit terms of $O(k)$ to Low's formula (\ref{Low}). The clue for the
solution is the observation that both $\cL_4$ and $\cL_4^{\Delta S=1}$
give rise to at most three independent four--derivative couplings at
the mesonic level:
$$
D_\mu \vp_a D^\mu \vp_b D_\nu \vp_c D^\nu \vp_d, \qquad
D_\mu \vp_a D^\mu \vp_c D_\nu \vp_b D^\nu \vp_d, \qquad
D_\mu \vp_a D^\mu \vp_d D_\nu \vp_b D^\nu \vp_c
$$
\beq
D_\mu \vp_a = (\partial_\mu + i q_a e A_\mu)\vp_a~.
\eeq
At the same time, we have three independent second derivatives of the
non--radiative amplitude $A(s,\nu)$. Therefore, the following extension of
(\ref{Low}) solves the problem 
(GB stands for ``generalized bremsstrahlung"):
\beqa
\label{EGB}
E^\mu &=& E^\mu_{\rm GB} + O(k) \no \\
E^\mu_{\rm GB} &=& A(s,\nu) \Sigma^\mu + 
2 \frac{\partial A(s,\nu)}{\partial s} \Lambda^\mu_{12} + 
\frac{\partial A(s,\nu)}{\partial \nu}(\Lambda^\mu_{13} - \Lambda^\mu_{23})
\no \\
&& \mbox{} +
2 \frac{\partial^2 A(s,\nu)}{\partial s^2} (t_1 + t_2)\Lambda^\mu_{12} + 
\frac{1}{2}
\frac{\partial^2 A(s,\nu)}{\partial \nu^2} [(t_1 - t_2)
(\Lambda^\mu_{13} - \Lambda^\mu_{23}) - t_3 t_4 \Sigma^\mu] \no \\
&& \mbox{} + 2 \frac{\partial^2 A(s,\nu)}{\partial s \partial \nu}
[t_2 \Lambda^\mu_{13} - t_1 \Lambda^\mu_{23}]~.
\eeqa

It is important to realize that $E^\mu_{\rm GB}$ in (\ref{EGB}) does not
contain {\bf all} terms of at most $O(k)$. In fact, it is impossible
in general to relate all terms of $O(k)$ to derivatives of $A(s,\nu)$.
On the other hand, the definition of generalized
bremsstrahlung in (\ref{EGB}) guarantees that all local (counter)terms
that contribute to both radiative and non--radiative processes (classes
A,B,C) are included in $E^\mu_{\rm GB}$. The difference 
$E^\mu - E^\mu_{\rm GB}$ is at least $O(k)$ and will be referred
to as the direct emission amplitude (of the electric type).

\paragraph{4.} We now turn to the loop amplitude. Most of the
renormalization procedure can trivially be carried over from the
non--radiative to the radiative amplitude because all diagrams 
(tadpoles) relevant for mass, charge and wave function renormalization
contribute only to internal bremsstrahlung. Thus, this part is completely
taken care of by the non--radiative amplitude. Moreover, for a real
photon there are no diagrams of the form factor type where the photon
emerges from a mesonic bubble.

Restricting attention to transitions of at most first order in the
Fermi coupling constant, the only non--trivial diagram 
is of the type shown in Fig.~1 where the photon can hook on
to any charged meson line and to any vertex with at least two charged
fields. The two vertices are either both from the lowest--order chiral
Lagrangian $\cL_2$ \cite{GL84,GL85} (strong transition) or one from
$\cL_2$ and one from $\cL_2^{\Delta S=1}$ (cf., e.g., 
Ref.~\cite{DEIN95}) for a nonleptonic weak transition. Despite the
comparative simplicity, diagrams of the type displayed in Fig.~1 with
a photon in all possible places generate a considerable number of
terms due to the derivative structure of vertices. Moreover, there
are usually several permutations $(1234) \ra (abcd)$ that have to be
added for a given process.

\begin{figure}
\centerline{\epsfig{file=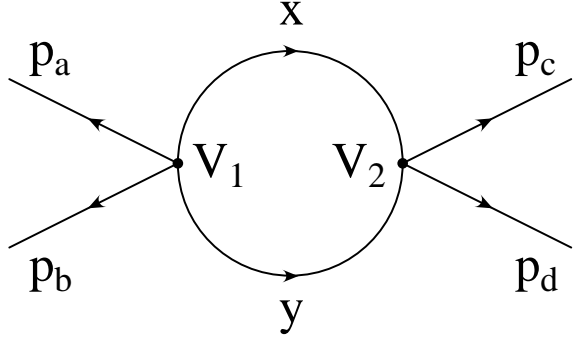,height=4cm}}
\caption{One--loop diagram for the four--meson transition.
For the radiative amplitude, the photon must be appended to every
charged meson line and to every vertex with at least two charged fields.
The vertices $V_1,V_2$ are defined in Eq.~(\protect{\ref{vertex}}).}
\end{figure}

We have therefore found it useful, both for our own work and for possible
future applications, to present the loop amplitude in a compact form
suitable for any strong or nonleptonic weak transition. For this purpose,
we first calculate the loop contribution to the non--radiative amplitude
$A(s,\nu)$.

We characterize the vertices $V_1,V_2$ in momentum space 
by real constants $a_i,b_i$:
\beqa
\label{vertex}
V_1 &=& a_0 + a_1 p_a \czdot p_b + a_2 p_a \czdot x + a_3(x^2 - M_x^2)
+ a_4(y^2 - M_y^2) + a_5(p_a^2 - M_a^2) + a_6(p_b^2 - M_b^2) \no \\
V_2 &=& b_0 + b_1 p_c \czdot p_d + b_2 p_c \czdot x + b_3(x^2 - M_x^2)
+ b_4(y^2 - M_y^2) + b_5(p_c^2 - M_c^2) + b_6(p_d^2 - M_d^2)~.\no \\
\eeqa
In calculating the loop amplitude for the non--radiative process, we do
not associate the various scalar products with the Dalitz variables $s,\nu$ 
that will depend on the specific assignment $(abcd) \ra (1234)$. Instead, 
we write the loop amplitude as a function of the four--momentum
\beq
P = p_c + p_d~.
\eeq
The non--photonic loop amplitude of Fig.~1 can be represented in the
following form that will turn out to be useful (all external lines
are on--shell):
\beqa
\label{Floop}
F(P) &=& A(M_x) [a_1 b_4 p_a \czdot p_b + a_4 b_1 p_c \czdot p_d +
a_4 b_4 (P^2 + M_x^2 - M_y^2) + a_0 b_4 + a_4 b_0] \no \\
&& \mbox{} + A(M_y) [a_1 b_3 p_a \czdot p_b + a_2 b_3 p_a \czdot P +
a_3 b_1 p_c \czdot p_d + a_3 b_2 p_c \czdot P \no \\
&& \hspace*{1.8cm} + a_3 b_3(P^2 - M_x^2 + M_y^2) + a_0 b_3 + a_3 b_0] 
\no \\
&& \mbox{} + B(P^2,M_x,M_y)[a_0 b_0 + a_0 b_1 p_c \czdot p_d +
a_1 b_0 p_a \czdot p_b + a_1 b_1 p_a \czdot p_b p_c \czdot p_d] \no \\
&& \mbox{} + B_1(P^2,M_x,M_y)[a_0 b_2 p_c \czdot P + a_2 b_0 p_a \czdot P
+ a_1 b_2 p_a \czdot p_b p_c \czdot P + a_2 b_1 p_c \czdot p_d p_a \czdot P] 
\no \\
&& \mbox{} + a_2 b_2[p_a \czdot p_c B_{20}(P^2,M_x,M_y) + p_a \czdot P
p_c \czdot P B_{22}(P^2,M_x,M_y)]~.
\eeqa
The various functions in (\ref{Floop}) are as defined conventionally
(in $d$ dimensions):
\beqa
\label{ABdef}
A(M) &=& \frac{1}{i} \int \frac{d^d x}{(2\pi)^d} 
\frac{1}{x^2 - M^2} \no \\
(B,B_1 P_\mu,g_{\mu\nu} B_{20} + P_\mu P_\nu B_{22}) &=&
\frac{1}{i} \int \frac{d^d x}{(2\pi)^d}
\frac{(1,x_\mu,x_\mu x_\nu)}{(x^2 - M_x^2)[(x-P)^2 - M_y^2]}~.
\eeqa
Several comments are in order at this point.
\begin{enumerate}
\item[i.] In order to limit the number of terms generated,
it is preferable to express the functions $B_1$,
$B_{20}$ and $B_{22}$ via the usual recursion 
relations in terms of $A,B$ only at the very end.
\item[ii.] The analytically non--trivial part of (\ref{Floop}), involving
the various $B$ functions, contains only the on--shell couplings
$a_0,a_1,a_2,b_0,b_1,b_2$. The off--shell couplings $a_3,a_4,b_3,b_4$
appear only together with the divergent constants $A(M)$. Since
these terms are polynomials in the momenta of at most degree two,
they will enter in the radiative amplitude only through internal
bremsstrahlung. All the divergences in (\ref{Floop}) will be absorbed
by counterterms belonging to classes A,B,C of the previous
classification. As emphasized before, the generalized bremsstrahlung
(\ref{EGB}) contains all these divergences plus the corresponding
counterterms.
\item[iii.] For a reason soon to become evident, we have chosen to
express $F(P)$ in terms of the scalar products
\beq
\label{scalar}
p_a \czdot p_b, \quad p_c \czdot p_d, \quad P^2, \quad
p_a \czdot P, \quad p_c \czdot P, \quad p_a \czdot p_c~.
\eeq
In other words, we have not used kinematical relations to write $F(P)$
in terms of only two independent scalar variables [like $s,\nu$
defined in Eq.~(\ref{kinema})].
\end{enumerate}

\paragraph{5.} We now have all the ingredients for calculating the 
radiative loop amplitude $E^\mu_{\rm loop}$ corresponding to the diagrams
of Fig.~1 with a photon in all possible places. For the general vertices
$V_1,V_2$ given in (\ref{vertex}), this loop amplitude contains several
hundred terms even before reducing the various $B$ functions via
recursion relations. A compact representation will therefore be of great
use for avoiding tedious repetitions of the same procedure.

We find it useful to decompose the radiative loop amplitude into two
parts:
\beq
E^\mu_{\rm loop} = G^\mu + H^\mu . \label{decomp}
\eeq
The more tedious part of the calculation is contained in the amplitude
$G^\mu$ that can be expressed through various derivatives of the
non--radiative loop amplitude $F$ in (\ref{Floop}) with respect to the
scalar products (\ref{scalar}). In some of the following terms, the
momentum $P$ has to be replaced by $P + k$, leaving all scalar products
unchanged that do not contain $P$ explicitly:
\beqa
\label{Gmu}
G^\mu &=& F(P) \Sigma^\mu + \frac{F(P+k) - F(P)}{k \czdot P} \Lambda^\mu_{cd}
+ \frac{\partial F}{\partial(p_a \czdot p_b)}(P) \Lambda^\mu_{ab} \no \\
&& \mbox{} + \frac{\partial F}{\partial(p_a \czdot P)}(P)\Lambda^\mu_{aP} +
\frac{\partial F}{\partial(p_c \czdot p_d)}(P+k)\Lambda^\mu_{cd} 
+ \frac{\partial F}{\partial (p_c \czdot P)}(P+k)\Lambda^\mu_{cP} \no \\
&& \mbox{} + \left[ q_a t_c \frac{\partial F}{\partial(p_a \czdot p_c)}(P)
- q_c t_a \frac{\partial F}{\partial(p_a \czdot p_c)}(P+k)\right]
D^\mu_{ac} \no \\
&& \mbox{} - \frac{1}{2}(q_c + q_d) t_a t_c \left[ 
\frac{\partial^2 F}{\partial(p_a \czdot P)\partial(p_c \czdot P)}(P)
D^\mu_{aP} - \frac{\partial^2 F}{\partial(p_a \czdot P)\partial(p_c \czdot
P)}(P+k) D^\mu_{cP} \right].\no \\
\eeqa
We have used the definitions (\ref{defs}). When $P$ appears as an index
(e.g., in $\Lambda^\mu_{aP}$ or $D^\mu_{cP}$), the corresponding momentum 
and charge in (\ref{defs}) are $P$ and $q_c + q_d$, respectively. For
better understanding of the notation in (\ref{Gmu}), we give two
explicit examples:
\beqa
\frac{\partial F}{\partial(p_c \czdot P)}(P+k) &=&
a_3 b_2 A(M_y) + (a_0 b_2 + a_1 b_2 p_a \czdot p_b)
B_1((P+k)^2,M_x,M_y) \no \\*
&& \mbox{} + a_2 b_2 p_a \czdot (P+k) B_{22}((P+k)^2,M_x,M_y) \no \\
\frac{\partial^2 F}{\partial(p_a \czdot P)\partial(p_c \czdot P)}(P)
&=& a_2 b_2 B_{22}(P^2,M_x,M_y)~.
\eeqa

The second part $H^\mu$ of the loop amplitude (\ref{decomp}) cannot
be expressed in terms of $F$ or derivatives thereof. Since the dominant
contributions to $E^\mu_{\rm loop}$ are usually due to pion loops (if
they contribute at all), we give the explicit expression for $H^\mu$
only for the case of equal loop masses ($M_x = M_y =: M$). In this
special case, $H^\mu$ takes on the following compact form:
\beqa
\label{Hmu}
H^\mu &=& a_2(t_b p^\mu_a - t_a p^\mu_b) \{(q_x-q_y)(2b_0 + 2b_1 p_c \czdot
p_d +b_2 p_c \czdot P) \wt{C_{20}}(P^2, - k \czdot P) \no \\
&& \mbox{} + b_2(q_x+q_y) [-2p_c \czdot P
\wt{C_{31}}(P^2,-k \czdot P) + 2t_c \wt{C_{32}}(P^2,-k \czdot P) 
- p_c \czdot P \wt{C_{20}}(P^2,-k \czdot P)]\} \no \\
&& \mbox{} + b_2(t_d p^\mu_c - t_c p^\mu_d) \{(q_x - q_y)
[2a_0 + 2a_1 p_a \czdot p_b + a_2(p_a \czdot P + t_a)] 
\wt{C_{20}}((P+k)^2, k \czdot P) \no \\
&& \mbox{} + a_2(q_x + q_y) [-2(p_a \czdot P + t_a)
\wt{C_{31}}((P+k)^2,k \czdot P)  - 2t_a \wt{C_{32}}((P+k)^2,
k \czdot P)\no \\ 
&& \mbox{} - (p_a \czdot P + t_a) \wt{C_{20}}((P+k)^2,k \czdot P)]\}~.
\eeqa
The functions $\wt{C_{ij}}$ are defined as
\beq
\wt{C_{ij}}(u,v) = \frac{C_{ij}(u,v) - C_{ij}(u,0)}{v} \label{Cdiff}
\eeq
in terms of the three--propagator one--loop functions 
$C_{ij}(p^2,k \czdot p)$ for $k^2 = 0$:
\beqa
\label{Cdef}
\lefteqn{ \frac{1}{i} \int \frac{d^dx}{(2\pi)^d} \frac{ \{x_\mu x_\nu,
x_\mu x_\nu x_\rho\}}{(x^2 - M^2)[(x+p)^2 - M^2][(x+k)^2 - M^2]} = } \no \\
&=& \{ C_{20}(p^2,k\czdot p)g_{\mu\nu} + \ldots, C_{31}(p^2,k \czdot p)
(p_\mu g_{\nu\rho} + p_\nu g_{\mu\rho} + p_\rho g_{\mu\nu}) \no \\
&& \mbox{} + C_{32}(p^2,k \czdot p)(k_\mu g_{\nu\rho} + k_\nu g_{\mu\rho}
+ k_\rho g_{\mu\nu}) + \ldots\}.
\eeqa
As in the case of $F(P)$ in (\ref{Floop}), it is advisable not to
use the standard recursion
relations for the functions $C_{20}$, $C_{31}$ and $C_{32}$ in
(\ref{Hmu}) until the actual numerical analysis. At the expense of
introducing the functions $B$, $B_{20}$, $B_{22}$ defined in (\ref{ABdef}),
one may express the functions $\wt{C_{ij}}((P+k)^2,k \czdot P)$ in
terms of the $\wt{C_{ij}}(P^2,- k \czdot P)$ or vice versa. 

The following comments (valid also in the case of 
different loop masses) explain the motivation for splitting the loop
amplitude $E^\mu_{\rm loop}$ in (\ref{decomp}) into two parts.
\begin{enumerate}
\item[i.] The amplitudes $G^\mu$ in (\ref{Gmu}) and $H^\mu$ in
(\ref{Hmu}) are separately gauge invariant.
\item[ii.] The amplitude $H^\mu$ is finite and at least of $O(k)$ as
is evident from Eqs.~(\ref{Hmu}), (\ref{Cdiff}) and (\ref{Cdef}).
Moreover, it only contains the on--shell couplings
$a_0,a_1,a_2,b_0,b_1,b_2$ defined in (\ref{vertex}) and the charges
$q_x,q_y$ of the particles in the loop. Of course, we have
\beq
q_x + q_y = - q_a - q_b = q_c + q_d~.
\eeq
\item[iii.] The amplitude $G^\mu$ contains the generalized bremsstrahlung
(\ref{EGB}) for the non--radiative loop amplitude (\ref{Floop}).
If we denote by  $E^\mu_{\rm GB}({\rm loop})$  the result obtained by
inserting for $A(s,\nu)$ the on--shell loop amplitude (\ref{Floop})
in Eq.~(\ref{EGB}), then the difference
\beq
\Delta^\mu = G^\mu - E^\mu_{\rm GB}({\rm loop})  
\eeq
is at least of $O(k)$. Moreover, by construction of $E^\mu_{\rm GB}$
the divergences in $\Delta^\mu$ are renormalized by counterterms of
class D only, i.e. by counterterms with an explicit field strength
tensor. In the strong sector, the relevant couplings of $O(p^4)$ are $L_9$ 
for chiral $SU(3)$ [$\ell_6$ for $SU(2)$] and $N_{14},N_{15},N_{16},N_{17}$
for the octet part of the nonleptonic weak Lagrangian \cite{EKW93}.
Finally, it can be shown that if $a_2 b_2 = 0$ then  
$\Delta_\mu$ is finite and at least of $O(k^2)$ 
for $s=\sum_{i=1}^{4} M_i^2/3$, $\nu=0$ and arbitrary $t_i$.
                                       
\item[iv.] The apparent asymmetry of $G^\mu$ and $H^\mu$ under
interchanges $a \leftrightarrow b$ or $c \leftrightarrow d$ is due
to the asymmetric definition of vertices (\ref{vertex}). For the same
reason, $G^\mu$ and $H^\mu$ are in general not invariant under
interchanges of the loop particles $x$ and $y$.
\end{enumerate}

For the realistic case with experimental information on the non--radiative
amplitude $A(s,\nu)$, the complete electric amplitude to $O(p^4)$
accuracy can be written as
\beq
E^\mu = E^\mu_{\rm GB}({\rm exp}) + E^\mu_{\rm counter}
+ \sum_{loops} (\Delta^\mu +H^\mu),  \label{Efin}
\eeq
where several loop diagrams may have to be added for a given physical
transition. Only counterterms with an explicit field strength tensor
must be included in $E^\mu_{\rm counter}$. Consequently, only the
renormalized coupling constants $L^r_9(\ell^r_6)$ and/or
$N^r_{14},\ldots,N^r_{17}$ appear in (\ref{Efin}). Of course, the
amplitude is finite and scale independent by construction. All other
counterterms of $O(p^4)$ are hidden in $E^\mu_{\rm GB}$(exp). An
alternative approach is to use instead of the experimental amplitude
$E^\mu_{\rm GB}$(exp) in (\ref{Efin}) the theoretical prediction
$E^\mu_{\rm GB}$(theory) in terms of the amplitude $A(s,\nu)$ calculated
in CHPT to $O(p^4)$ accuracy. Both approaches are equivalent to 
$O(p^4)$. The difference between them gives an indication of the
size of effects of $O(p^6)$ and higher.

\paragraph{6.} As a final application, we consider the limit $q_a \ra 0$,
$p_a \ra 0$ to connect with known results for $K \ra 2\pi\gamma$ decays.
In this case, the vertex $V_1$ is necessarily of the weak nonleptonic
type and the coupling constants $a_1,a_2,a_5$ disappear. A straightforward
calculation shows that $G^\mu$ becomes in this limit
\beq
G^\mu = F(P+k) \Sigma^\mu \label{brems}
\eeq
where $F(P+k) = F(-p_b)$ is now the on--shell loop amplitude for the decay
$b \ra c + d$. Eq.~(\ref{brems}) is nothing but the familiar 
bremsstrahlung amplitude for a radiative three--meson transition and
as a consequence
\beq
\Delta^\mu=0~.
\eeq
Likewise, the amplitude $H^\mu$ in (\ref{Hmu}) reduces
in the three--meson limit to the single term \cite{DMS93,ENP94,DI95}
\beq
H^\mu = 2a_0 b_2(q_x - q_y) \wt{C_{20}} (M^2_b,-k \czdot p_b)
(t_d p_c^\mu - t_c p^\mu_d).
\eeq
Thus, in this limit, the loop contribution to the direct emission 
is finite and proportional to $a_0$, the on--shell tree--level 
amplitude for the nonleptonic weak transition $b \ra c + d$.
                                                                  
\paragraph{7.} To summarize, we have presented a general discussion of
radiative four--meson processes to $O(p^4)$ in CHPT. We have proposed
a definition of generalized bremsstrahlung in Eq.~(\ref{EGB}) that has
the advantage of including all counterterms
of $O(p^4)$ that contribute to both radiative and non--radiative 
amplitudes. For general vertices of $O(p^2)$ that encompass all strong
and nonleptonic weak transitions of interest, we have calculated the
non--trivial loop amplitudes in terms of two gauge invariant parts.
The amplitude $G^\mu$ given in (\ref{Gmu}) is expressed in terms of
the non--radiative loop amplitude $F$ in (\ref{Floop}). The divergences
are all contained in $G^\mu$. The remainder $H^\mu$ given in
Eq.~(\ref{Hmu}) is finite and at least of $O(k)$.

Applications of these general results to $K \ra 3\pi\gamma$ decays and 
other radiative four--meson processes will be presented elsewhere
\cite{DEIN962}.

%\newpage
%\begin{center}
%{\large \bf Acknowledgements}
% \end{center} 
%\noindent
 
%\newpage

\end{document}